\documentclass[11pt,a4paper]{article}
\usepackage{hyperref}
\usepackage{bm}
\usepackage{amsmath} 
\usepackage[utf8]{inputenc} 
\usepackage{amsthm,amssymb,lmodern}

\usepackage{cite}

\usepackage{array}
\usepackage{cancel}
\newcommand\ie{\textit{i.e.}\ }
\newcommand\eg{\textit{e.g.}\ }

\newcommand{\be}{\begin{equation}}
\newcommand{\ee}{\end{equation}}
\newcommand{\bea}{\begin{eqnarray}}
\newcommand{\eea}{\end{eqnarray}}

\newcommand{\gz}{g^{(0)}}
\newcommand{\tz}{t^{(0)}}
\newcommand{\Tz}{T^{(0)}}
\newcommand{\z}{{(0)}}
\newcommand{\weak}{\ \underset{weak}{=}\ }
\newcommand{\G}{{\cal G}}

\begin{document}
\begin{titlepage}
\begin{flushright}
\end{flushright}

\begin{center}
{\huge \bf Cosmological back-reaction  in modified gravity
and its implications for 
dark energy}

\end{center}
\vskip1cm


\begin{center}
{\bf Anthony W. H. Preston \& Tim R. Morris}
\end{center}

\begin{center}
{\it School of Physics and Astronomy,  University of Southampton\\
Highfield, Southampton, SO17 1BJ, U.K.}\\
\vspace*{0.3cm}
{\tt  T.R.Morris@soton.ac.uk, awhp1g12@soton.ac.uk}
\end{center}


%

\abstract{
We study the effective stress-energy tensor induced by cosmological inhomogeneity in  $f(R)=R+c R^2$ and equivalent scalar-tensor theories, motivated both by models of early universe inflation and by phenomenological alternative cosmologies to the standard $\Lambda$-CDM. We use Green and Wald's 
framework for averaging over classical fluctuations of short-wavelength $\lambda$. By ensuring that the leading non-linear terms  from the fluctuations of the Einstein terms and the corrections both contribute in the formal limit as $\lambda\to0$, we derive a diffeomorphism invariant effective stress-energy tensor whose trace is non-vanishing and of the right sign to 
potentially account for the current acceleration of the universe. However a more phenomenologically acceptable dark energy model would be required if this effect were to fully account for the current acceleration.}



\end{titlepage}

\section{Introduction}

The standard cosmological model describes our universe using a Friedmann-Letma\^\i tre-Robertson-Walker (FLRW) metric. Underlying this description is the Cosmological Principle, the assumption that the mass density in the universe, on sufficiently large scales, is to good approximation homogeneous and isotropic. On small scales, the mass distribution is patently far from uniformly distributed. For example the ratio of the matter density on Earth to the average mass-energy density in the universe is $\delta\rho/\rho\sim10^{31}$, or for individual nucleons  $\delta\rho/\rho\sim10^{46}$.  But in the standard cosmological model the assumption is that  for deriving the dynamics of the metric on the largest scales, these variations can effectively be replaced by some large scale average. 

However, since Einstein's equations of General Relativity (GR)  are non-linear, this averaging process does not result in Einstein's equations for some `average' metric,  but alters the equations themselves. Many papers have been written trying to determine if this so-called ``cosmological back-reaction'' could thus be responsible for the measured current acceleration of the universe (see for example refs. \cite{Rasanen:2003fy,Schwarz:2010px,Clarkson:2011zq,Clifton:2013vxa,Roukema:2013cya,Buchert:1995fz,Buchert:1999er,Roy:2011za,Buchert:2011sx}), rather than the result of an effective cosmological constant arising from some new and mysterious ``dark energy'' physics or some surprisingly small net positive vacuum energy density. 

Furthermore since sufficiently large amplitude inhomogeneities have appeared only recently in the history of the universe, cosmological back-reaction has the potential to provide a natural explanation for the infamous ``Why now?'' problem, namely that it seems that energy densities of dark energy and dark matter are now, but only now, similar in magnitude.

Of course such a scenario would then imply that the underlying (quantum field theoretic) net vacuum energy actually vanishes, which itself has no theoretical explanation. Nevertheless it is clearly important  to understand the  extent to which back-reaction contributes to the measured current acceleration of the universe, in order to be able to isolate and characterise any remaining dark energy component. It is this latter component which would then apparently require some more fundamental theoretical explanation.

We will be interested in one particularly  elegant approach to the question of back-reaction, by Green and Wald \cite{Green:2010qy,Green:2013yua}. As we will explain in detail later,
they perform the averaging process rigorously in a weak limit where mass density fluctuations (and hence also higher derivatives of the metric) can remain arbitrarily large but the metric itself tends to some smooth background. 
Assuming that the matter stress-energy tensor $T_{\alpha \beta}$ satisfies the weak energy condition, Green and Wald prove that 
the averaged effect of the coupled matter plus gravitational fluctuations is then encoded in  this limit in an additive correction $\tz_{\alpha \beta}$ 
to the stress-energy which is traceless and also satisfies the weak energy condition. They therefore identify it 
with gravitational radiation.
In particular 
in a FLRW background metric, $\tz_{\alpha\beta}$ is diagonal, corresponding to an effective fluid with pressure $p=\rho/3\ge0$, leading to the conclusion that this back-reaction cannot mimic dark energy. 

In the current paper we will generalise some of their arguments to $f(R)$ gravity and scalar-tensor theory. In fact for $f(R)$ we will take only Einstein gravity plus an $R^2$ correction:
\begin{equation}
\label{action0}
 S = \int d^{4}x\sqrt{-g}\left[\frac{1}{16\pi G}
 \left(R+ 
 \frac{R^2}{6M^2} -2\Lambda
 \right)+\mathcal{L}_{Matter}\right]\,.
\end{equation} 
We set the coefficient $c$ quoted in the abstract, to $c=1/6M^2$, where $M$ is known as the scalaron mass \cite{Starobinsky:1980te}. Although it plays no r\^ole in the paper, we keep a cosmological constant $\Lambda$ for the sake of generality.

After appropriate modifications of the Green and Wald scheme, we find that back-reaction is again encoded in a diffeomorphism invariant effective additive correction $\tz_{\alpha\beta}$ to the matter stress energy tensor, however \emph{it is now  not traceless}. 
In fact
the sign of $\tz$ is fixed to be negative, intriguingly the same sign as implied by the current acceleration of the universe. 

In order to gain further confidence in our result and to generalise it, we then treat the equivalent scalar-tensor theory in a similar way. We find the same diffeomorphism invariant effective stress-energy tensor $\tz_{\alpha\beta}$ with negative-definite trace, and  we can provide some more physical intuition as to why this is the case. 

The action \eqref{action0} was first introduced with the motivation to include semi-classical quantum effects; it is the Starobinsky model of ``$R^2$ inflation'' \cite{Starobinsky:1980te,Davies:1977ze}, one of the earliest models of inflation,  and one which is  favoured observationally by WMAP \cite{Hinshaw:2012aka} and particularly the  Planck measurements of CMB anisotropy \cite{Ade:2013uln}. The prediction of only a tiny amount of primordial gravitational waves puts this model now in tension with the recent BICEP2 results \cite{Ade:2014xna}, but these are in tension with the results of the Planck collaboration (for further discussion see \eg \cite{Smith:2014kka,Li:2014cka}) and some questions have been raised about their foreground estimates \cite{Liu:2014mpa,Flauger:2014qra}.

It is not our intention to enter the debate over 
these recent results however, but rather to use the Starobinsky model as the simplest archetype of a motivated extended gravity theory and to ask the question again within this context whether cosmological back-reaction could contribute significantly to, or even be entirely responsible for, the measured current acceleration of the universe. 

In fact the same action \eqref{action0} appears in quite a different context, namely as the simplest example of $f(R)$ gravity regarded as a purely phenomenological effective Lagrangian density describing geometrical dark energy and  used in the quest to find an alternative cosmology to the standard $\Lambda$-CDM model \cite{Hu:2007nk,Starobinsky:2007hu,Tsujikawa:2007xu,Appleby:2007vb,Linder:2009jz,Amendola:2006we,Li:2007xn,DeFelice:2010aj,Tsujikawa:2010sc,Abebe:2013zua,Clifton:2011jh,Guo:2013swa,Mukherjee:2014fna}. We can then regard \eqref{action0} as following from the leading terms in a Taylor expansion of an $f(R)$ which is analytic around the vacuum solution $R=0$. Although such a simple choice as \eqref{action0} is excluded, we believe our results for back-reaction should motivate a reassessment of these theories in general. 



There is some overlap with studies already in the literature. 
Refs. \cite{Vitagliano:2009zy,Jimenez:2013mwa}
are studies that are rather close in spirit to ours but which follow Buchert's averaging scheme \cite{Buchert:1995fz,Buchert:1999er}. Using this scheme, they argued that even in cases where back-reaction is negligible for GR with matter, back-reaction might indeed give rise to non-negligible effects for modified cosmologies. However, tentative arguments in support of some contribution to dark energy have also been made for just GR and matter using such averaging schemes, see \eg refs. \cite{Roy:2011za,Buchert:2011sx}. Green and Wald criticised such approximate approaches and conclusions in ref. \cite{Green:2010qy}, contrasting with their own negative conclusions regarding the form of $\tz_{\alpha\beta}$. In ref. \cite{Buchert:2011sx}, Green and Wald's scheme was in turn criticised on the grounds that it assumes from the beginning that the metric remains close to the FLRW background. 
In any case, we therefore think it is especially significant that even staying within the Green and Wald approach
we find a negative definite $\tz$ in our example of modified cosmology.

The Green and Wald scheme is itself a generalisation to the non-vacuum case of the framework used by Burnett \cite{Burnett:1989gp}  in his mathematically rigorous formulation of Isaacson's shortwave approximation to gravitational waves \cite{Isaacson:1967zz,Isaacson:1968zza}. 
Papers that use Isaacson's scheme directly to investigate modified gravity theories are refs. \cite{Stein:2010pn,Berry:2011pb,Saito:2012xa}. Refs. \cite{Stein:2010pn,Berry:2011pb} consider gravitational waves on a Minkowski background to place observational constraints on modified gravity theories such as $f(R)$ theories, and derive a corresponding effective stress energy tensor in Lorenz gauge. Saito and Ishibashi \cite{Saito:2012xa} work in a cosmological context but find, in contrast to our own conclusions, that the effective stress-energy tensor is still traceless in such $f(R)$ and scalar-tensor theories. 
The crucial reasons for the disagreement are that 
they explicitly ignore fluctuations in the matter fields, and only average over gravitational fluctuations, whereas we require that the back-reaction has a non-vanishing weak limit from averaging over coupled fluctuations in both matter and gravitational fields. To go into this further and other technical differences we postpone the rest of the comparison to the conclusions.

In the conclusions, we emphasise that while our key technical assumption in the averaging procedure is that the leading non-linear terms from both the Einstein part and the $R^2$ part remain finite and contribute in the weak limit $\lambda\to0$, nevertheless 
the effective stress-energy tensor gives sensible results in the limits where only one of these parts dominate. Thus we show that for inhomogeneity length scales $L$ that are much longer than the scalaron scale: $L\gg1/M$, the effective stress-energy tensor $\tz_{\alpha\beta}$ reduces to the pure Einstein case \cite{Green:2010qy} as required. 

Also in the conclusions, we discuss further the physical context and implications of our result. One would need to establish if for sensible choices of $1/M$, $\tz$ could actually be of the right magnitude and also be approximately constant, if it is to be the sole cause the current acceleration of the universe. As we discuss, this seems unlikely for such a simple model as \eqref{action0}, but we hope that the results of our analysis motivate a reassessment of such modified gravity models in general to take into account these back-reaction effects.

The structure of the paper is as follows. In the next section, we set out in detail our technical assumptions, and in particular in what sense we generalise those of Green and Wald  \cite{Green:2010qy}. In sec. \ref{calculation} we compute the effective stress-energy tensor and show that its trace part is negative definite. In sec. \ref{diffeomorphism} we show that the result is diffeomorphism invariant. In sec. \ref{scalar-tensor} we derive the result in the equivalent scalar-tensor theory, and finally in sec. \ref{conclusions} we draw our conclusions.

\section{The setup}

We adopt the Landau-Lifshitz spacelike sign conventions (+,+,+) such that the metric has signature $(-+++)$, the Ricci tensor $R_{\mu\nu} = R^{\alpha}_{\ \mu \alpha \nu}$, and
\be
R^{\mu}_{\ \nu\rho\sigma} = 2\,\partial_{[\rho}\Gamma^{\mu}_{\ \sigma]\nu} + 2 \,\Gamma^{\mu}_{\ \lambda[\rho} \Gamma^{\lambda}_{\ \sigma]\nu}\,,
\ee
where the Levi-Civita connection is defined in the usual way
\begin{equation}
 \Gamma^{\mu}_{\ \nu\rho} = \frac{1}{2}g^{\mu \alpha}(\partial_{\nu} g_{\rho \alpha} + \partial_{\rho} g_{\nu \alpha} - \partial_{\alpha} g_{\nu \rho})\,.
\end{equation}
Varying \eqref{action0} with respect to the metric, the field equations are
\be
\label{eom0}
\G_{\mu\nu} = \kappa T_{\mu\nu}\,,
\ee
where $\kappa := 8\pi G$ and the Einstein tensor plus Starobinsky corrections are given by 
\begin{equation}
\label{eom}
 \G_{\mu\nu} := R_{\mu\nu}+\Lambda g_{\mu\nu} -\frac{1}{2}g_{\mu\nu}\left(R+\frac{R^2}{6M^2}\right) +\frac{1}{3M^2}\left(RR_{\mu\nu}-D_{\mu}D_{\nu}R+g_{\mu\nu}\,D^2 R\right)\,,
\end{equation}
where $D_\mu$ is the covariant derivative, and  $D^2 :=g^{\mu\nu}D_\mu D_\nu$.


While the huge variations in mass density in our own universe imply  that there are huge 
variations in space-time derivatives of the metric, we expect that the deviation of the metric itself from FLRW form remains small almost everywhere. In order to model this, we adapt the mathematical conditions imposed in  ref. \cite{Green:2010qy},\footnote{We try to cast the analysis in somewhat less mathematical language, in particular we eschew the explicit use of abstract space-time indices which we could have used throughout. We trust so-inclined readers can put these details back for themselves.} which are themselves generalisations to the non-vacuum case of the conditions used by Burnett \cite{Burnett:1989gp}  in his mathematically rigorous formulation of Isaacson's shortwave approximation to gravitational waves \cite{Isaacson:1967zz,Isaacson:1968zza}.

For reasons that will become clear later we regard the scalaron mass as dependent on the inhomogeneity parameter $M\equiv M(\lambda)$. We can make this generalisation because the r\^ole of $\lambda$ is only to parametrise in a coordinate-invariant way a family of solutions to \eqref{eom} chosen by us
such that the metric $g_{\mu\nu}(\lambda,x)$ and matter stress-energy tensor $T_{\mu\nu}(\lambda,x)$ 
are jointly smooth in $\lambda$ and space-time coordinates $x$. 
Properties of the inhomogeneity will effectively be parameterised by $\lambda>0$. 
The metric is taken to converge uniformly (on compact sets) as $\lambda\to0$ to a ``background metric'' $\gz_{\mu\nu}(x) :=g_{\mu\nu}(0,x)$, thus 
encapsulating the notion 
that the magnitude of the fluctuations in the metric $h_{\mu\nu}(\lambda,x) := g_{\mu\nu}(\lambda,x)- \gz_{\mu\nu}(x)$ remain everywhere small (for sufficiently small $\lambda$). In fact we further require that  $h_{\mu\nu}$ is $O(\lambda)$.\footnote{For our purposes, this is equivalent to the condition stated in ref. \cite{Green:2010qy}. Throughout the paper, by a tensor being $O(\lambda^n)$  strictly we  mean that for sufficiently small $\lambda$, and on compact sets, its components can be bounded uniformly by a constant times $\lambda^n$.} Note that during our analysis we do not however assume any special solution for $g^{(0)}_{\mu\nu}$,  in particular we do not assume FLRW form. 

The matter is taken to satisfy the weak energy condition, \ie for all $\lambda>0$ we have 
\be
\label{weak-energy}
T_{\alpha\beta}(\lambda,x)t^\alpha(\lambda,x)t^\beta(\lambda,x)\ge0
\ee 
for all vector fields $t^\alpha(\lambda,x)$ which are time-like with respect to $g_{\mu\nu}(\lambda,x)$. Unlike the metric, the stress-energy tensor is allowed to fluctuate wildly in the sense that it will not converge in the limit $\lambda\to0$.

Thus through the field equations \eqref{eom},  space-time derivatives of $h_{\mu\nu}$ will also not converge in the limit $\lambda\to0$. Introducing the background covariant derivative $\nabla_\mu$, we can nevertheless impose that first derivatives $\nabla_\alpha h_{\mu\nu}$ are $O(1)$ (\ie uniformly bounded on compact sets). A simple one-dimensional example is 
\be
\label{example}
h\sim \lambda\sin(x/\lambda)\,.
\ee

Now let $f^{\alpha\beta\gamma}$ be any smooth tensor field of compact support, then 
\be
\int\! d^{4}x\sqrt{-\gz} f^{\alpha\beta\gamma} \,\nabla_\alpha h_{\beta\gamma} = - \int\! d^{4}x\sqrt{-\gz} \,\nabla_\alpha f^{\alpha\beta\gamma}\,  h_{\beta\gamma} \to 0\quad{\rm as}\quad \lambda\to0\,.
\ee
This integral amounts to averaging over the finite amplitude, but infinitesimally small distance features that remain in $\nabla_\alpha h_{\mu\nu}$ as $\lambda\to0$. Since the result is zero for any such test tensor field $f^{\alpha\beta\gamma}$, we say that $\nabla_\alpha h_{\beta\gamma}$ converges weakly to zero and write:
\be
\label{diff-h}
\nabla_\alpha h_{\beta\gamma} \weak 0\,.
\ee
More generally a tensor $A_{\mu\nu\cdots}$  converges weakly to its `average' tensor $B_{\mu\nu\cdots}$, which we express as $A_{\mu\nu\cdots}\weak B_{\mu\nu\cdots}$, if
\begin{equation}
 \lim_{\lambda\rightarrow 0}\int \! d^{4}x\sqrt{-\gz} f^{\alpha\beta\cdots}A_{\alpha\beta\cdots}(\lambda) = \int \! d^{4}x\sqrt{-\gz} f^{\alpha\beta\cdots}B_{\alpha\beta\cdots}
\end{equation}
for any smooth $f^{\mu\nu\cdots}$ of compact support. 

If $C_{\mu\nu\cdots}$ is another tensor that converges weakly to the same average tensor, then as we will see, we will find it useful to write
\be
\label{weak-relation}
A_{\mu\nu\cdots}\weak C_{\mu\nu\cdots}\,, 
\ee
where we then effectively regard the average tensor as defined by the weak limit of $C_{\mu\nu\cdots}$.

The final mathematical condition we take from ref. \cite{Green:2010qy} is that the weak limit of 
\be
\label{mu}
\nabla_\alpha h_{\mu\nu} \nabla_\beta h_{\rho\sigma}
\ee 
exists, this being a smooth tensor field $\mu_{\alpha\beta\mu\nu\rho\sigma}$. It is such terms that are responsible for giving the effective stress energy tensor found by Green and Wald. These set of conditions allow us to isolate 
the leading non-trivial back-reaction coming from $h_{\mu\nu}$ in Einstein's part of the equations \eqref{eom}, within a mathematically precise framework. Note that as a consequence of all these conditions, 
\be
\nabla_\alpha \left(h_{\mu\nu} \nabla_\beta h_{\rho\sigma}\right)\weak 0\,, 
\ee
since after integrating by parts the total covariant derivative onto the test tensor field the remaining parts are $O(\lambda)$. This allows us to place both covariant derivatives on either $h_{\alpha\beta}$ (keeping track of signs of course). Furthermore they can be placed in either order since the commutator $[\nabla_\alpha,\nabla_\beta]$ will just yield background curvature terms, leaving terms of $O(\lambda^2)$.

The Starobinsky corrections to \eqref{eom} however have the potential to spoil all this since after weak averaging we are left with terms of the form 
\be
\label{four-derivs}
{1\over M^2} \nabla_\alpha\nabla_\beta h_{\mu\nu} \nabla_\gamma\nabla_\delta h_{\rho\sigma}\,,
\ee
which we can expect to be unbounded as $\lambda\to0$. 
It is for this reason we take $M$ to depend on $\lambda$. We then require that \eqref{four-derivs} is $O(1)$ and that its weak limit exists and is a smooth tensor field.\footnote{For example using  \eqref{example}, we would achieve this with $M=m/\lambda$ for some constant $m$. Note that as in ref. \cite{Green:2010qy} we are making the presumably mild assumption that such families of solutions exist. This was further explored in \cite{Green:2013yua}. The only changes to the requirements on our set of solutions are the new equations of motion (\ref{eom0},\ref{eom}) and this extra condition.}  We similarly see that in these terms we have the freedom to place the covariant derivatives on either $h_{\alpha\beta}$ and in any order.

In this way we will isolate in a mathematically controlled way the leading non-trivial back-reaction 
coming from the both Einstein's part of the equation \eqref{eom} and the Starobinsky corrections together. 
The difference 
\be
\label{difference}
\delta\left[ \G_{\mu\nu} \right] = \G_{\mu\nu} - \G^\z_{\mu\nu}
\ee
(where here and later the superscript $^{(0)}$ indicates that $g_{\alpha\beta}$ is replaced by the background metric), is actually unbounded in the limit $\lambda\to0$. However the problem lies only in the linearised parts which on weak averaging vanish, similarly to \eqref{diff-h}.
Thus we will find that \eqref{difference} has a non-vanishing weak limit, which we view as an effective correction to the stress-energy:
\be
\label{effective}
\delta\left[ \G_{\mu\nu} \right]  \weak -\kappa\, \tz_{\mu\nu}\,.
\ee
The equation of motion which we now write as
\be
\label{diff-eom}
\G^\z_{\mu\nu} + \delta\left[ \G_{\mu\nu} \right] = \kappa T_{\mu\nu}\,,
\ee
holds for all $\lambda$. We see that $T_{\mu\nu}$ is therefore also unbounded in the limit $\lambda\to0$. But since by assumption the left hand side has a weak limit, we have as in ref. \cite{Green:2010qy}, that $T_{\mu\nu}$ must also have a weak limit $\Tz_{\mu\nu}$. Crucial use will be made of the assumption that $T_{\mu\nu}$ satisfies the  weak energy condition since the property \eqref{weak-energy} is required to prove that
\be
\label{hT}
h_{\alpha\beta} T_{\mu\nu} \weak 0\,.
\ee
As in ref. \cite{Green:2010qy}, this follows from the lemma proved there that $A(\lambda)B(\lambda) \weak A(0) B(0)$ if $A$ is a smooth tensor field converging uniformly on compact sets to $A(0)$ and $B$ is a non-negative smooth function converging weakly to $B(0)$. 

We will raise and lower indices using $\gz_{\alpha\beta}$. Utilising the freedom to move background covariant derivatives about, it is convenient to define 
\be
\label{R4}
R^{(1)}_{\mu\alpha\nu\beta} := -2 \nabla_{[\mu|}\nabla_{[\nu} h_{\beta]\,|\alpha]}\,,
\ee
this being the linearised Riemann tensor. From this we have also the linearised Ricci tensor
\begin{equation}\label{R}
 R^{(1)}_{\mu\nu} :=\frac{1}{2}\left(2\nabla_{\alpha}\nabla_{(\mu}h_{\nu)}^{\ \ \alpha}-\Box h_{\mu\nu} - \nabla_{\mu}\nabla_{\nu} h\right)\,,
\end{equation}
where we define $\Box := g^{(0)\alpha\beta}\nabla_{\alpha}\nabla_{\beta}$ and $h=h^\alpha_{\ \ \alpha}$, and the linearised Ricci scalar
\be
R^{(1)} := \nabla^\alpha\nabla^\beta h_{\alpha\beta}-\Box h \,.
\ee
As will be seen in sec. \ref{diffeomorphism}, $R^{(1)}_{\mu\nu\alpha\beta}$ and its contractions are effectively diffeomorphism invariant when they appear in appropriate expressions in the weak limit.

\section{The calculation}
\label{calculation}

We have everything now in place to present the details of the calculation. The difference between the full connection and background connection is the tensor
\begin{equation}
  C^{\mu}_{\ \nu\rho} =  \frac{1}{2}g^{\mu \alpha}(\nabla_{\nu} h_{\rho \alpha} + \nabla_{\rho} h_{\nu \alpha} - \nabla_{\alpha} h_{\nu \rho})\,.
\end{equation}
Thus the Ricci tensor splits up as follows
\begin{equation}
 R_{\mu\nu} = R^{(0)}_{\mu\nu} - 2\nabla_{[\mu}C^{\alpha}_{\ \alpha]\nu} + 2C^{\alpha}_{\ \nu[\mu}C^{\beta}_{\ \beta]\alpha}.
\end{equation}
To the order in which we are working, we can consistently neglect all but the first power of $h_{\alpha\beta}$ in the inverse metric:
\begin{equation}
\label{higher}
 g^{\mu\nu}(\lambda,x) = g^{\mu\nu}(0,x)-h^{\mu\nu}(\lambda,x) + O(h^{2}).
\end{equation}
In \eqref{effective}, terms with only one $h_{\alpha\beta}$ have zero weak limit, leaving terms with two $h_{\alpha\beta}$ as the only contribution to the averaged back-reaction. We thus obtain the same intermediate\footnote{\ie before simplification using the Ricci form of Einstein's equations.} result from the Einstein parts of \eqref{effective} as Green and Wald \cite{Green:2010qy}:
\begin{eqnarray}
\kappa t_{\mu\nu}^{E} &=& \frac{1}{8}g_{\mu\nu}^{(0)}(-\mu^{\alpha\ \beta\gamma}_{\ \alpha \ \ \beta\gamma} - \mu^{\alpha\ \beta\ \gamma}_{\ \alpha\ \beta\ \gamma} + 2\mu^{\alpha\beta\ \gamma}_{\ \ \alpha\ \beta\gamma}) \nonumber \\ 
&& + \frac{1}{2}\mu^{\alpha\beta}_{\ \ \mu \alpha \nu \beta} -\frac{1}{2}\mu^{\alpha\ \ \beta}_{\ \alpha \mu \ \nu \beta} + \frac{1}{4}\mu_{\mu\nu \ \ \alpha\beta}^{\ \ \alpha\beta} \nonumber \\
&& -\frac{1}{2}\mu^{\alpha\ \ \ \ \ \beta}_{\ (\mu\nu)\alpha\ \beta} +\frac{3}{4}\mu^{\alpha \ \ \ \beta}_{\ \alpha\mu\nu\ \beta} -\frac{1}{2}\mu^{\alpha\beta}_{\ \ \mu\nu \alpha\beta}\,,
\end{eqnarray}
where recall $\mu_{\alpha\beta\mu\nu\rho\sigma}$ was introduced through \eqref{mu}. This can alternatively be written in a more convenient way for us as 
\be 
\label{Einstein}
\kappa t_{\mu\nu}^{E} \weak {1\over2} h^{\alpha\beta}R^{(1)}_{\mu\alpha\nu\beta} +{3\over4}h_{\mu\nu}R^{(1)}-R^{(1)}_{\alpha (\mu} h_{\nu)}^{\ \ \alpha} -{1\over8}\gz_{\mu\nu}\left( hR^{(1)} + 2 h^{\alpha\beta}R^{(1)}_{\alpha\beta} \right)
\ee
in the sense of \eqref{weak-relation}. Similarly, forming differences as in \eqref{difference}, the terms from the Starobinsky parts of the field equation that survive weak averaging  are
\bea
\label{Star RR}
{\delta\left[ R_{\mu\nu}R \right]/ M^2} &\weak& R^{(1)}_{\mu\nu}R^{(1)}/M^2\,, \\
\label{Star R2}
\delta\left[ g_{\mu\nu}R^{2}\right]/M^2 &\weak& g_{\mu\nu}^{(0)}{R^{(1)}}^2/M^2\,, \\
\label{Star DDR}
\delta \left[D_{\mu}D_{\nu}R \right]/M^2 &\weak& \frac{1}{2}\left(2R^{(1)}_{\mu\nu} +\nabla_{\mu}\nabla_{\nu}h\right)R^{(1)}/M^2\,, \\
\label{Star D2R}
\delta\left[ g_{\mu\nu}D^2 R \right]/M^2 &\weak& \left(\Box h_{\mu\nu}-\frac{1}{2}g^{(0)}_{\mu\nu}\Box h\right)R^{(1)}/M^2\,.
\eea
Combining and simplifying, we can write the averaged contribution to the back-reaction from the Starobinsky terms as
\begin{equation}\label{StaroNotZeroStress}
 \kappa t^{S}_{\mu\nu} \weak \frac{R^{(1)}}{3M^2}\left({1\over2}\gz_{\mu\nu}\Box h-\Box h_{\mu\nu} +\frac{1}{2}\nabla_{\mu}\nabla_{\nu}h  +{1\over4}\gz_{\mu\nu} R^{(1)}\right)\,.
\end{equation}

The weak energy condition, through \eqref{hT}, 
leads to a crucial constraint, which we refer to as a ``zero tensor''. This encodes information  from the total derivative single $h_{\alpha\beta}$ terms which have so far been discarded. The most general zero tensor can be obtained by first moving to the Ricci form. The trace of \eqref{eom} is 
\begin{equation}
\label{trace-eom}
4\Lambda -R+\frac{D^2 R}{M^2}= \kappa T,
\end{equation}
therefore we can eliminate either $R$ or ${D^2 R}/{M^2}$. These then yield two related Ricci forms, which in turn yield two related forms for the zero tensor upon multiplication by $h_{\rho\sigma}$ and weak averaging. These  are
\begin{equation}\label{zero1}
 h_{\rho\sigma}R^{(1)}_{\mu\nu}-\frac{g^{(0)}_{\mu\nu}}{6M^2}\Box h_{\rho\sigma}R^{(1)} - \frac{1}{3M^2}(\nabla_{\mu}\nabla_{\nu}h_{\rho\sigma})R^{(1)} \underset{weak}{=} 0
\end{equation}
and
\begin{equation}\label{zero2}
 h_{\rho\sigma}R^{(1)}_{\mu\nu}-\frac{g_{\mu\nu}^{(0)}}{6}h_{\rho\sigma}R^{(1)} - \frac{1}{3M^2}(\nabla_{\mu}\nabla_{\nu}h_{\rho\sigma})R^{(1)} \underset{weak}{=} 0.
\end{equation}
As with the Ricci forms, we can obtain one from the other by back substitution of the trace (over $\mu\nu$):
\be
\label{zero-diff}
R^{(1)} \left(1-{\Box\over M^2}\right) h_{\rho\sigma} \weak 0\,.
\ee
Up to a factor this is also the difference of the two equations \eqref{zero1} and \eqref{zero2}, and
of course it also follows directly from multiplying \eqref{trace-eom} by  $h_{\alpha\beta}$ and taking the weak limit.
Equation (\ref{zero1}) is good for converting 2-derivative expressions into 4-derivative expressions and (\ref{zero2}) is good for the opposite. Indeed all the ${O}(h^2)$ 4-derivative expressions we found can be converted into 2-derivative expressions using (\ref{zero2}). This exchange may seem counter-intuitive, but if the Einstein-Hilbert term existed in isolation, its contributions to the zero tensors would average to zero, similarly for the Starobinsky terms. 
In our case we have the terms from the two parts added together; neither of them average to zero separately but influence each other as should be expected.

We can use  (\ref{zero2}) now to rewrite (\ref{StaroNotZeroStress}) in 2-derivative form. Doing this, we get
\begin{equation}
\label{t-Starobinsky}
 \kappa t^{S}_{\mu\nu} \underset{weak}{=} \frac{1}{2}hR^{(1)}_{\mu\nu}+\frac{g^{(0)}_{\mu\nu}}{4}h^{\alpha\beta}R^{(1)}_{\alpha\beta}-\frac{1}{3}h_{\mu\nu}R^{(1)}-\frac{g^{(0)}_{\mu\nu}}{24}hR^{(1)}.
\end{equation}
We then add this to the contribution from the Einstein-Hilbert terms in equations (\ref{Einstein}) to get the overall stress-energy tensor in 2-derivative form, which is
\be
\label{OverallStress}
\kappa \tz_{\mu\nu}\weak {1\over2} h^{\alpha\beta}R^{(1)}_{\mu\alpha\nu\beta} -R^{(1)}_{\alpha (\mu} h_{\nu)}^{\ \ \alpha}+{5\over12}h_{\mu\nu}R^{(1)}+{1\over2}hR^{(1)}_{\mu\nu} -{1\over6}\gz_{\mu\nu}hR^{(1)} 
\ee
The trace of this is clearly
\begin{equation}
\label{trace-2d}
 \kappa t^{(0)} \underset{weak}{=} \frac{1}{4}hR^{(1)}-\frac{1}{2}h^{\alpha\beta}R^{(1)}_{\alpha\beta},
\end{equation}
which is almost trivially diffeomorphism invariant. The overall weak-averaged effective stress-energy tensor will be shown to be diffeomorphism invariant in the next section. We can convert the trace back into 4-derivative form using (\ref{zero1}) and get
\begin{equation}
\label{trace-4d}
 \kappa t^{(0)} \underset{weak}{=} -\frac{{R^{(1)}}^2}{6M^2}.
\end{equation}
This form will not only make the diffeomorphism invariance of the trace more obvious, but also shows that it is negative definite. 
We can split the effective stress-energy tensor into traceless and pure trace parts as follows
\begin{equation}
 t^{(0)}_{\mu\nu} = \underbrace{t^{(0)}_{\mu\nu} - \frac{1}{4}g^{(0)}_{\mu\nu}t^{(0)}}_{\text{traceless}} + \underbrace{\frac{1}{4}g^{(0)}_{\mu\nu}t^{(0)}}_{\text{pure trace}}.
\end{equation}
The pure trace term has the correct sign 
to mimic a positive cosmological constant, in accordance with observation of an accelerating universe.

\section{Diffeomorphism invariance}
\label{diffeomorphism}

For our result to be physically meaningful, it should also be diffeomorphism invariant. The infinitesimal diffeomorphism transformation can be taken to be the first order result in an expansion in $h_{\alpha\beta}$:
\begin{equation}
\label{diff}
 h_{\mu\nu} \rightarrow h_{\mu\nu} +2\nabla_{(\mu}\xi_{\nu)}.
\end{equation}
Strictly speaking we should write the vector field as $\xi_\alpha(x) = \epsilon\, {\tilde\xi}_\alpha(x)$, where $\epsilon$ is vanishingly small, and 
where for small $\lambda$, ${\tilde\xi}_\alpha$ is $O(\lambda^2)$ and $\nabla_\alpha {\tilde\xi}_\beta$ is $O(\lambda)$. It is these latter properties that allow us to restrict to the first order in the weak limit since we can consistently neglect the higher order corrections in $h_{\alpha\beta}$ to this expression when using it in weak limits, for similar reasons to the neglect of powers in \eqref{higher}. 

Within expressions that have a weak limit, and utilising the fact that background covariant derivatives then effectively commute, $R^{(1)}_{\mu\nu\alpha\beta}$ and thus also its traces, can be seen trivially to be invariant under the above transformation \eqref{diff}. It immediately follows that the trace \eqref{trace-4d} of the effective stress energy tensor is also invariant.
To show that the overall weak-averaged effective stress-energy tensor is diffeomorphism invariant, we must make use of a zero tensor. First, we take (\ref{zero2}) and antisymmetrise in the indices $[\mu,\rho]$ and $[\nu,\sigma]$ to get
\begin{equation}
 \frac{1}{6}g^{(0)}_{[\mu|[\nu}h_{\sigma]|\rho]}R^{(1)}-h_{[\rho|[\sigma}R^{(1)}_{\nu]|\mu]} \underset{weak}{=} -\frac{R^{(1)}}{3M^2}\nabla_{[\mu|}\nabla_{[\nu}h_{\sigma]|\rho]} = {1\over6M^2} R^{(1)} R^{(1)}_{\mu\rho\nu\sigma}.
\end{equation}
Under diffeomorphisms \eqref{diff}, the right-hand side is invariant in the weak limit, leaving
\begin{multline}
\nabla_{(\rho}\xi_{\sigma)}R^{(1)}_{\mu\nu} + \nabla_{(\mu}\xi_{\nu)}R^{(1)}_{\rho\sigma} - \nabla_{(\mu}\xi_{\sigma)}R^{(1)}_{\nu\rho} - \nabla_{(\nu}\xi_{\rho)}R^{(1)}_{\mu\sigma} \underset{weak}{=} \\ 
\frac{g^{(0)}_{\mu\nu}}{6}\nabla_{(\rho}\xi_{\sigma)}R^{(1)}+\frac{g^{(0)}_{\rho\sigma}}{6}\nabla_{(\mu}\xi_{\nu)}R^{(1)}-\frac{g^{(0)}_{\mu\sigma}}{6}\nabla_{(\nu}\xi_{\rho)}R^{(1)}-\frac{g^{(0)}_{\nu\rho}}{6}\nabla_{(\mu}\xi_{\sigma)}R^{(1)}.
\end{multline}
To create a useful rank two zero tensor from this, we contract $\rho$ and $\sigma$ using the background metric to get
\begin{equation}\label{zerogauge}
 \frac{g_{\mu\nu}^{(0)}}{6}\nabla_{\alpha}\xi^{\alpha}R^{(1)} \underset{weak}{=} \nabla_{\alpha}\xi^{\alpha}R^{(1)}_{\mu\nu}+\frac{2}{3}\nabla_{(\mu}\xi_{\nu)}R^{(1)} -\nabla_{(\mu|}\xi_{\alpha}R_{\ |\nu)}^{(1)\ \alpha}-\nabla_{\alpha}\xi_{(\mu}R_{\nu)}^{(1)\ \alpha}.
\end{equation}
The na\"{i}ve gauge transformation of (\ref{OverallStress}) is
 \begin{eqnarray}
\kappa\, \delta t^{(0)}_{\mu\nu} &\underset{weak}{=}& \frac{g_{\mu\nu}^{(0)}}{3}\xi^{\alpha}\nabla_{\alpha}R^{(1)} -\frac{1}{3}\xi_{(\mu}\nabla_{\nu)}R^{(1)} - \xi_{\alpha}\Box\nabla_{(\mu}h_{\nu)}^{\ \ \alpha} \nonumber \\
&& + \xi_{\alpha}\nabla_{\beta}\nabla_{\mu}\nabla_{\nu}h^{\alpha\beta} - \xi^{\alpha}\nabla_{\alpha}\nabla_{\beta}\nabla_{(\mu}h_{\nu)}^{\ \beta} + \xi_{\alpha}\nabla^{\alpha}\Box h_{\mu\nu}.
\end{eqnarray}
Substituting (\ref{zerogauge}) into the first term in the above equation and cancelling terms leaves $\delta t^{(0)}_{\mu\nu}$ equal to zero. Thus the effective stress-energy tensor is invariant under linearized diffeomorphisms using this averaging proceedure.

\section{Scalar-tensor theory}
\label{scalar-tensor}

An alternative description of $f(R)$ gravity is scalar-tensor gravity. By deriving the effective stress-energy tensor in both descriptions, we are able to check the consistency of our approach.
The action for scalar-tensor theory in Jordan frame, assuming the Brans-Dicke parameter to be zero, can be written as
\begin{equation}
\label{BDzeroAction}
S = \int d^{4}x\sqrt{-g}\left[\frac{1}{2\kappa}\left(\phi R - V(\phi) -2\Lambda \right) + \mathcal{L}_{matter}\right], 
\end{equation}
where $\phi$ is a dimensionless scalar. We set $V(\phi)$ to have its minimum at zero by separating out the cosmological constant from the potential. The metric field equation is
\begin{equation}
\label{ST-metric-eom}
 \phi \left(R_{\mu\nu} - \frac{1}{2}g_{\mu\nu}R \right)+ g_{\mu\nu}\Lambda  = \kappa T_{\mu\nu} + \left(D_{\mu}D_{\nu}-g_{\mu\nu}D^2\right)\phi-\frac{1}{2}g_{\mu\nu}V(\phi).
\end{equation}
Varying $\phi$ in \eqref{BDzeroAction} gives
$R=V'(\phi)$ which, combined with the trace of above, gives us the scalar field equation
\begin{equation}
\label{ST-scalar-eom}
 3D^2 \phi = \kappa T +\phi V'(\phi)-2V(\phi)-4\Lambda.
\end{equation}
The scalar-tensor theory is equivalent to $f(R)$ gravity under the Legendre transformation $f(R)=\phi R - V(\phi)$. 
Starting with $f(R) = R + \frac{1}{6M^2}R^2$, the scalar potential in scalar-tensor theory is then
\begin{equation}
\label{pot}
 V(\phi) = \frac{3M^2}{2}\left(\phi-1\right)^2.
\end{equation}
We approach averaging and perturbation theory in the same way as before, but now we include perturbations in the scalar field as well,
\begin{equation}
\label{scalar-exp}
 \phi (x, \lambda) = \phi_0(x) + \phi_{p}(x, \lambda),
\end{equation}
such that $\phi(x, \lambda)\to\phi_0(x)$ as $\lambda\to0$. In the Starobinsky case
\begin{equation}
 \phi = f'(R) = 1 + \frac{R}{3M^2}.
\end{equation}
Since we take $1/M$ to be $O(\lambda)$, we see that we have simply
\begin{equation}
\label{vev}
 \phi_{0} = 1, 
\end{equation}
and
\begin{equation}
\label{pert}
 \phi_{p}(x, \lambda) = \frac{R}{3M^2}\,, 
\end{equation}
where $\phi_p$ is $O(\lambda)$. We will call $\phi_{0}$ the background field and $\phi_{p}$ the perturbation, however note that from \eqref{pert}, $\phi_p$ has a piece also dependent on the background curvature $R^{(0)}$ which however is forced to vanish as a result of $M(\lambda)\to\infty$.

Consistently with \eqref{pert} we require that $\nabla_\mu\phi_p$ is $O(1)$. From \eqref{four-derivs} we already have that 
\be
\label{phi-h}
\phi_{p}\nabla_{\mu}\nabla_{\nu}h_{\rho\sigma}
\ee
has a weak limit which is a smooth tensor field. Substituting \eqref{scalar-exp}, \eqref{vev} and \eqref{pert}, into \eqref{pot}, we see that $V(\phi)$ is $O(1)$ and has a weak limit which is a smooth non-negative scalar field, which we name $V_p(x)$. Recalling that the matter stress-energy tensor is precisely the same as the one previously, we have shown that everything in the metric field equation \eqref{ST-metric-eom}  has a well-defined weak limit. 

In the general case, using a general positive semi-definite potential $V(\phi)$, it is then natural to assume that $V(\phi)$ and terms such as \eqref{phi-h} have a weak limit. Similarly to the derivation of \eqref{hT}, we then deduce that $h_{\rho\sigma}V(\phi)$ has vanishing weak limit. In \eqref{ST-metric-eom} we have then ensured that everything apart from $T_{\mu\nu}$ has a weak limit. As below \eqref{diff-eom} this allows us to deduce that $T_{\mu\nu}$ also has a weak limit. Similarly, from the scalar field equation \eqref{ST-scalar-eom} we then see that $\phi V'(\phi)$ also has a weak limit; in the Starobinsky case, using \eqref{scalar-exp}, \eqref{vev}, \eqref{pert}, and \eqref{pot}, this is of course clear directly.

When constructing zero tensors, we follow the same procedure as in the $f(R)$ case. The trace of the metric field equation is
  \begin{equation}
    -\phi R = \kappa T -3D^2 \phi -2V -4\Lambda,
  \end{equation}
therefore one can construct Ricci forms by using this to remove either $\phi R$ or $\Box \phi$ from the metric field equation. One does not gain any new information by using a Ricci form obtained from removing $V(\phi)$, since $h_{\rho\sigma}V(\phi)$ is vanishing in weak limit anyway. 
The two related forms of the zero tensor obtained from the metric field equation are
\begin{equation}
\label{scalar-tensor-zero-1}
 0 \underset{weak}{=} h_{\rho\sigma}R^{(1)}_{\mu\nu} -\frac{1}{2}g^{(0)}_{\mu\nu}h_{\rho\sigma}\Box\phi_{p}-h_{\rho\sigma}\nabla_{\mu}\nabla_{\nu}\phi_{p},
\end{equation}
\begin{equation}
\label{scalar-tensor-zero-2}
 0 \underset{weak}{=} h_{\rho\sigma}R^{(1)}_{\mu\nu} - \frac{1}{6}g^{(0)}_{\mu\nu}h_{\rho\sigma}R^{(1)} - h_{\rho\sigma}\nabla_{\mu}\nabla_{\nu}\phi_{p}.
\end{equation}
and correspond to those in the $f(R)$  description. This is readily seen in the Starobinsky case by using \eqref{pert} to substitute the leading piece, $\frac{R^{(1)}}{3M^2}$, for $\phi_{p}$. We see immediately that equations (\ref{scalar-tensor-zero-1}) and (\ref{zero1}) are the same and (\ref{scalar-tensor-zero-2}) and (\ref{zero2}) are also the same.

Although it is less useful for our purposes, one can also construct a zero tensor from the scalar field equation, which is
\begin{equation}
 0 \underset{weak}{=} 3h_{\rho\sigma}\Box\phi_{p} -h_{\rho\sigma}V'(\phi).
\end{equation}
If we suppose that there also exist weak limits for terms of the forms $\phi_{p}\nabla_{\mu}\nabla_{\nu}\phi_{p}$ and $\phi_{p}V'(\phi)$, then one can also construct zero tensors where $h_{\rho\sigma}$ is replaced with $\phi_{p}$. These zero tensors are also less useful for our purposes.



Now we consider the weak limit of the metric field equation \eqref{ST-metric-eom} and compare it to the $f(R)$ description. The terms from the left hand side with $\phi$ replaced by $\phi_0=1$ have the same weak limit as in (\ref{Einstein}), the remaining perturbation pieces of the field equation have the following weak limits
\bea
\label{ST RR}
\phi_{p}R_{\mu\nu} &\underset{weak}{=}& \phi_{p}R^{(1)}_{\mu\nu},\\
\label{ST R2}
\phi_{p}g_{\mu\nu}R &\underset{weak}{=}& g^{(0)}_{\mu\nu}R^{(1)}\phi_{p},\\
\label{ST V}
\delta\left[g_{\mu\nu}V\right] &\underset{weak}{=}& g^{(0)}_{\mu\nu}V_{p} \weak \frac{3}{2}g^{(0)}_{\mu\nu}M^{2}\phi_{p}^{2},\\
\label{ST DDR}
\delta\left[D_{\mu}D_{\nu}\phi\right]  &\underset{weak}{=}& \frac{1}{2}\left(R^{(1)}_{\mu\nu}+\nabla_{\mu}\nabla_{\nu}h\right)\phi_{p},\\
\label{ST D2R}
\delta\left[g_{\mu\nu}D^{2}\phi\right] &\underset{weak}{=}& \left(\Box h_{\mu\nu} - \frac{1}{2}g_{\mu\nu}^{(0)}\Box h\right)\phi_{p}.
\eea
Again, using the leading part of (\ref{pert}), it is easy to relate these to the corresponding terms in the $f(R)$ description for the Starobinsky case. In particular, equations (\ref{Star RR}) and (\ref{ST RR}) are the same, (\ref{Star R2}), (\ref{ST R2}) and (\ref{ST V}) are all proportional to each other in the weak limit, (\ref{Star DDR}) is the same as (\ref{ST DDR}), and finally (\ref{Star D2R}) is the same as (\ref{ST D2R}). The resulting effective stress-energy tensor can be expressed as (\ref{OverallStress}) using (\ref{scalar-tensor-zero-2}) to convert all terms into the Einstein 2-derivative form, except for the potential, 
\begin{eqnarray}
\kappa t^{(0)}_{\mu\nu} &\underset{weak}{=}& \frac{1}{2}h^{\alpha\beta}R^{(1)}_{\mu\alpha\nu\beta}-R^{(1)}_{\alpha(\mu}h_{\nu)}^{\ \alpha} + \frac{5}{12}h_{\mu\nu}R^{(1)}  +\frac{1}{2}hR^{(1)}_{\mu\nu}\nonumber\\ && +\frac{1}{4}g^{(0)}_{\mu\nu}h_{\alpha\beta}R^{(1)\alpha\beta}-\frac{7}{24}g_{\mu\nu}^{(0)}hR^{(1)} -\frac{1}{2}g_{\mu\nu}^{(0)}V_{p}.
\end{eqnarray}
(The potential term can however be turned into the 4-derivative form by (\ref{pert}) and then converted to a two-derivative form using (\ref{zero2}).)
The contribution in weak limit to $\kappa t^{(0)}$ from the potential perturbation is $-\frac{1}{3M^2}{R^{(1)}}^2$, the remaining ``kinetic terms" supply a correction of $+\frac{1}{6M^2}{R^{(1)}}^2$. We get the correct sign because the larger contribution is from the potential, whose minimum is set at zero, and the average perturbation to that potential is therefore necessarily above zero.

\section{Conclusions}
\label{conclusions}

As we reviewed in the introduction, Green and Wald were able to show that in General Relativity 
the effects of cosmological back-reaction could be summarised in a certain weak limit as an additive effective stress-energy tensor $\tz_{\alpha\beta}$ which was traceless and thus could not mimic dark energy \cite{Green:2010qy}. The weak limit is a rigorous  averaging scheme in which a family of solutions are dependent on some external parameter $\lambda$. This parameter can be thought of as representing the `inhomogeneity wavelength', and the limit that is taken is $\lambda\to0$. In this limit the metric is required to tend  to a smooth background (FLRW in this cosmological context) but higher space-time derivatives of the metric and the matter density contrast are still allowed to have divergent fluctuations. Terms such as \eqref{mu} however have a weak limit and are in fact responsible for the effective stress-energy tensor.

We have generalised this scheme to the action \eqref{action0} where Einstein gravity now has an $R^2$ correction, and the equivalent scalar tensor theory. A crucial step in this generalisation is that now both terms \eqref{mu} and  \eqref{four-derivs} are required to have a  weak limit that is {\it a priori} non-vanishing. This is necessary to ensure that the leading back-reaction is captured from both the Einstein and Starobinsky terms in the limit $\lambda\to0$. It follows that in our family of solutions, 
$M$ is also required to depend on $\lambda$.

In the weak limit the back-reaction is now summarised in the diffeomorphism invariant effective stress energy tensor $\tz_{\alpha\beta}$  given in \eqref{OverallStress}. As in ref. \cite{Green:2010qy}, we make crucial use here of the assumption that the matter stress-energy tensor satisfies the weak energy condition. This allows us to derive the 
 `zero tensor' (\ref{zero1}--\ref{zero-diff}) which further constrains the back-reaction. In particular the trace $\tz$ can then be written in two-derivative form as \eqref{trace-2d} or equivalently written in four-derivative form as \eqref{trace-4d}. The latter form is manifestly negative, the right sign to explain the current acceleration of the universe. We discuss this physical context later.

Although we kept the leading non-linearities from fluctuations in both the Einstein and Starobinsky terms, the resulting effective stress-energy tensor has the expected limit when only the Einstein term makes a significant contribution. This corresponds to the case where the inhomogeneity length scale $L\gg1/M$, or what is physically equivalent, the scalaron mass $M\to\infty$ whilst keeping (covariant) derivatives of $h_{\alpha\beta}$ finite. In this case the 
zero tensor collapses to the condition
\be
\label{GWzero}
h_{\rho\sigma} R^{(1)}_{\mu\nu} \weak 0,
\ee
which is the condition derived in ref. \cite{Green:2010qy} from assuming that the matter stress-energy tensor satisfies the weak energy condition. Applying \eqref{GWzero} to our effective stress-energy tensor \eqref{OverallStress} we see immediately that we recover the pure Einstein case: $\tz_{\mu\nu} \weak {1\over2} h^{\alpha\beta}R^{(1)}_{\mu\alpha\nu\beta}$, this being the effective stress-energy tensor obtained in ref. \cite{Green:2010qy} after simplification using \eqref{GWzero} (which in turn is the same as in ref. \cite{Burnett:1989gp}). From \eqref{trace-4d} we see immediately that we recover its tracelessness, as is also clear directly from \eqref{GWzero}.

In the opposite limit in which $L\ll1/M$, one might have expected the four-derivative terms from \eqref{StaroNotZeroStress} to dominate. However, were we able to neglect the Einstein parts entirely 
and thus retain only terms containing 
$1/M^2$, it is easy to see that the zero tensor
in particular form \eqref{zero2}, would imply that this contribution \eqref{StaroNotZeroStress} vanishes. 
In the rigorous framework we derive this as follows. The limit $L\ll 1/M$ corresponds to letting $\lambda\to0$ at fixed $M$. Since by assumption the field equations \eqref{eom0} are obeyed and have a weak limit, terms of the form \eqref{four-derivs} then force the vanishingly small wavelength fluctuations of the matter-gravity coupled system to have smaller amplitude. A simple one-dimensional example would be $h\sim \lambda^2 \sin(x/\lambda)$, which should be compared to \eqref{example}. Thus we verify that the pure Einstein terms, whose leading behaviour comes from terms \eqref{mu}, are forced to vanish in this limit whilst the contribution \eqref{StaroNotZeroStress} {\it a priori} survives. However only the $1/M^2$ piece of the zero tensor \eqref{zero2} survives and thus we see that in fact the effective stress energy tensor then vanishes completely.\footnote{Note that this is a consequence of imposing the weak energy condition, via \eqref{weak-relation}.} 

More generally, for any finite solutions such as might model the real universe (\ie not now taking the limit $\lambda\to0$) we see that the zero tensor implies that the contribution from the $R^2$ term grows only as $\sim1/L^2$ as follows from the equivalent formulation in 
\eqref{t-Starobinsky}. Although \eqref{trace-4d} suggests that the trace $\tz$ grows as $\sim1/L^4$, the zero tensor constraints ensure that it is equal to the two-derivative form \eqref{trace-2d} and thus only grows as $\sim1/L^2$.

Now we turn to a detailed comparison with the derivation of Saito and Ishibashi 
\cite{Saito:2012xa}. 
Using Isaacson's scheme \cite{Isaacson:1967zz,Isaacson:1968zza}, they classify contributions from fluctuations according to order, which in our notation is
\be
\label{orders}
\nabla_{\mu_1} \cdots \nabla_{\mu_n}h_{\alpha\beta}\sim O(\lambda^{1-n}).
\ee
Except for the fact that they take $M$ to be independent of $\lambda$, this is consistent with our assumptions. However they also ``for simplicity''  ignore perturbations in the matter fields in \eqref{eom0}, \emph{replacing} $T_{\alpha\beta}$  with the background matter stress-energy tensor $\Tz_{\alpha\beta}$. The leading contributions from gravitational fluctuations in \eqref{eom} are divergent $\sim O(\lambda^{-m})$, with $m=1,2,3$. Since these are however not sourced directly by matter, they are able to use the symmetries of the FLRW metric to solve iteratively the equations of motion for these pieces, starting with the most divergent. After a choice of integration constant, and averaging, they find their traceless effective stress-energy tensor at $O(\lambda^0)$. 

In contrast, we do allow both the matter stress-energy tensor $T_{\mu\nu}$ and the gravitational fields to fluctuate 
and since we also want to extract a non-vanishing weak limit as the leading contribution for the effective $\tz_{\mu\nu}$, we are then essentially forced to different conclusions. Recalling the discussion surrounding \eqref{hT}, and as in ref. \cite{Green:2010qy}, the crucial steps that allow us nevertheless to make progress are the proof that $T_{\mu\nu}$ has a weak limit  $\Tz_{\mu\nu}$   and that, providing $T_{\mu\nu}$ satisfies the weak energy condition, the product $h_{\alpha\beta} T_{\mu\nu}$ vanishes in the weak limit. Since our gravitational fluctuations are sourced by matter fluctuations (and of course also {\it vice versa}) we cannot  directly solve for the divergent $O(\lambda^{-m})$ pieces like in ref. \cite{Saito:2012xa}. Instead, by multiplying by an extra factor of $h_{\alpha\beta}$ the most divergent piece provides the four-derivative contributions to our zero tensors. On their own these pieces of the zero tensors are sufficient to force the leading $O(\lambda^{-2})$ pieces of the effective stress energy tensor to vanish, 
as we have seen already in the  discussion above about the regime $L\ll1/M$.  We would then be left with sub-leading $O(1/\lambda)$ pieces (containing four derivatives and three fluctuation fields $h_{\alpha\beta})$. In ref. \cite{Saito:2012xa} these can also be solved for, but since we have fluctuating matter, we need some analogue of \eqref{hT} to apply to sub-leading pieces. The difficulty is that $O(1/\lambda)$ pieces of the stress-energy tensor do not separately satisfy the weak energy condition so there is no such straightforward analogue of \eqref{hT}. 
Instead, we maintain mathematical control and also extract the leading contributions from both the Einstein and Starobinsky pieces, by effectively requiring also that $M\sim O(1/\lambda)$ in the formal limit $\lambda\to0$ as we have already discussed.

Finally, we discuss the physical context and physical implications of our result. From our discussion above, it is clear that inhomogeneities at length scales $L\lesssim1/M$ would make the most significant contribution. Intuitively the reason for this and the non-vanishing trace $\tz$ is the presence of the extra scalar (scalaron) mode with mass $M$ in such a theory. Although we have seen that the trace $\tz$ has the right sign, we would also need to establish that $\tz$ has the right magnitude and is also approximately constant in time, if it is to be the sole cause of the current acceleration of the universe. 

If we regard \eqref{action0} in its original incarnation as a model for inflation, then the scalaron mass is determined as  $M\approx 3\times10^{13}$ GeV, implicating  phenomena related to the Grand Unified Theory (GUT) scale of $\sim 10^{16}$ GeV. Example candidates might include perturbations sourced by WIMPzilla-like dark matter \cite{Kolb:1998ki,Chung:2001cb} whose high mass would introduce large spacetime derivatives close to the particle. One can speculate that, coupled to high frequency gravitational radiation also induced by back-reaction, this might induce suitably large average scalar perturbations. Even if one were prepared to go along with our speculations, it seems unlikely however that such phenomena would lead to a $\tz$ that is approximately constant in time, although for GUT-scale models with implications for cosmology with scalar-modes, this back-reaction would nevertheless need to be considered as part of the predictions. An even wilder speculation is that quantum fluctuations in spacetime might average to small classical perturbations at the GUT scale, whose back-reaction could be described with this formalism and might also be expected to be constant in time in the present epoch. However it is hard to ignore in these circumstances the infamous problem of the quantum fluctuations themselves which would naturally be expected  to give already a cosmological constant at the scale of the Planck mass. (Nor is it understood in the effective field theory language why the value of $M$ is so much smaller than the Planck mass.) 

On the other hand in the incarnation of \eqref{action0} as the simplest example of a phenomenological geometrical dark energy model, we can potentially explore much larger length scales $1/M$. In order to obtain the correct value for dark energy we can expect perhaps $1/M\sim 100$ Mpc, corresponding to the largest scale of inhomogeneities in the present universe. We might then also reasonably expect that such $\tz$ is approximately constant at the present time as a result of a competition between cosmological dilution of matter density and increasing amplitude of inhomogeneity. Unfortunately solar system and laboratory bounds exclude such large values of $1/M$ by many orders of magnitude \cite{Berry:2011pb,DeFelice:2010aj} and of course \eqref{action0} is also too simple to be a satisfactory dark energy model for cosmological evolution. Nevertheless we hope that our results will motivate a re-evaluation of such dark energy models in general 
\cite{Hu:2007nk,Starobinsky:2007hu,Tsujikawa:2007xu,Appleby:2007vb,Cognola:2007zu,Linder:2009jz,Amendola:2006we,Li:2007xn,DeFelice:2010aj,Tsujikawa:2010sc,Nojiri:2010wj,Abebe:2013zua,Clifton:2011jh,Guo:2013swa} to take into account the back-reaction effects studied in this paper.

\section{Acknowledgments}
TRM acknowledges STFC support through Consolidated Grant ST/J000396/1 and AWHP acknowledges support from the University of Southampton through a Mayflower scholarship. We thank  Christopher Berry, Cliff Burgess, Shaun Hotchkiss, Patrick Morris and Ippocratis Saltas for discussions.

\bibliographystyle{hunsrt}
\bibliography{refs}

\end{document}